\documentclass[prb,twocolumn,amsmath,amssymb,floatfix]{revtex4}
\usepackage{epsfig,color}
\usepackage{bm}

\usepackage{graphicx}

\newcommand{\be}{\begin{eqnarray}}
\newcommand{\ee}{\end{eqnarray}}
\newcommand{\bfr}{{\bf r}}

\newcommand{\wbe}{\begin{widetext}}
\newcommand{\wee}{\end{widetext}}
\newcommand{\oncite}{\onlinecite}

\begin{document}

\title{Quantum Criticality from {\it in-situ} Density Imaging}

\author{Shiang Fang$^{1}$, Chia-Ming Chung$^{1}$, Ping Nang Ma$^{2}$, Pochung Chen$^{1,3}$, Daw-Wei Wang$^{1,3}$}

\affiliation{
$^{1}$ Physics Department, National Tsing-Hua University, Hsinchu 300, 
Taiwan
\\
$^{2}$ Theoretische Physik, ETH Zurich, 8093 Zurich, Switzerland
\\
$^{3}$ Physics Division, National Center for Theoretical Sciences,
Hsinchu 300, Taiwan}

\date{\today}

\begin{abstract}
We perform large-scale Quantum Monte Carlo (QMC) simulations for strongly interacting bosons in a 2D optical lattice trap, and confirm an excellent agreement with the benchmarking {\it in-situ} density measurements by the Chicago group [\oncite{Chin}]. We further present a general finite temperature phase diagram both for the uniform and the trapped systems, and demonstrate how the universal scaling properties near the superfluid(SF)-to-Mott insulator(MI) 
transition can be observed by analysing the {\it in-situ} density profile. The characteristic temperature to find such quantum criticality is estimated to be of the order of the single-particle bandwidth, which should be achievable in the present or near future experiments. Finally, we examine the validity regime of the local fluctuation-dissipation theorem (FDT), which can be a used as a thermometry in the strongly interacting regime.
\end{abstract}


\maketitle

In this recent decade, systems of ultracold atoms have become the most promising candidate for the realization of quantum simulators, which are designed to explore various challenging and exotic many-body physics. One of the most well-understood ultracold systems is made of bosonic atoms loaded in an optical lattice, whereby the 3D SF-MI phase transition had been observed in time-of-flight(TOF) experiments[\oncite{Greiner}] and then directly compared to theory with great accuracy [\oncite{ETH}].
However, the TOF images do {\it not} directly represent physical quantities of atoms inside the optical lattice and thus subject the quantitative study of critical phenomena to question, especially given the complexity of the expansion dynamics during the TOF imaging [\oncite{Bloch,Fang}]. As a result, 
several schemes have been proposed to determine the critical properties based on density-related quantities[\oncite{Jason,Mueller,Lode_Criticality}], which have been successfully measured via {\it in-situ} spatially resolved density imaging in a bosonic systems loaded into a 2D optical lattice [\oncite{Chin,Greiner2D}]. Before these theoretical schemes could be trusted and applied, an essential step is to validate them from the first principle calculation in the parameter regime of realistic experiments.

\begin{figure}
\includegraphics[width=8.4cm]{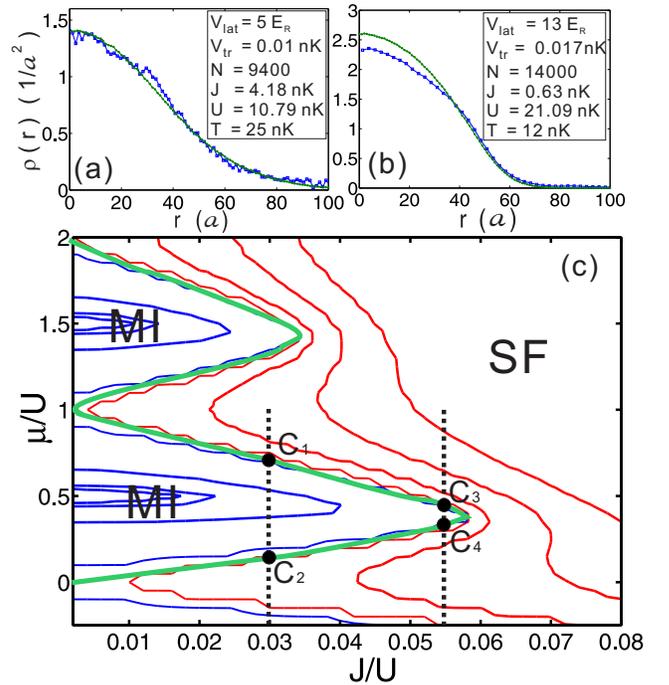}
\caption{(Color online) Density profiles measured by Chicago group [\oncite{Chin}] (blue squares) and by our QMC simulation (green lines) for (a) a shallow lattice and (b) a deep lattice. $a(V_{\rm lat})$ is the lattice constant(strength), and $E_R$ is the recoil energy. (c) is the finite temperature phase diagram of a uniform 2D system. Green thick line is the SF-MI phase boundary at $T=0$. Red(blue) lines are boundaries of SF(MI) at $T/U=0.01$, $0.05$. $0.1$, and $0.15$ from left to right(from right to left). The regime of MI phase is defined in the text. The vertical dashed lines are the chemical potential range in Figs. \ref{scaling} and \ref{ZEROT}.}
\label{CCexp}
\end{figure}
  
In this paper, we perform an {\it ab-initio} QMC simulation [\oncite{DWALODE}] to have a direct comparison with the {\it in-situ} data of Chicago group [\oncite{Chin}] in a 2D optical lattice with a harmonic trap, and find an excellent agreement (see Figs. \ref{CCexp}(a) and (b)). We then calculate the finite temperature phase diagram for a uniform system, and justify the validity of the local density approximation (LDA), except near the phase boundary. We further quantitatively confirm the universal scaling properties near the SF-MI quantum phase transition point through the finite temperature density profile in a trapped system. The characteristic temperature, $T^\ast$, to find such quantum criticality is comparable to the single particle band width, and therefore is achievable in the present or near future experiments. Finally, we show the regime where the local fluctuation-dissipation theorem (FDT) is valid, providing a useful thermometry in the strongly interacting regime. Our work therefore paves the way for the future experimental realization of quantum simulators.

Our numerical simulation starts from the single-band Bose-Hubbard model:
$\hat{H}= -J \sum_{\langle i,j\rangle}\left(\hat{a}_i^{\dagger} \hat{a}_j +{\rm h.c.}\right)+ \frac{U}{2}\sum_i \hat{n}_i (\hat{n}_i -1)- \sum_i \mu(\bfr_i)\hat{n}_i$, where $\hat{a}_i$ and $\hat{n}_i$ are the boson field operator and density operator at the $i$th lattice site.  $\mu(\bfr_i)\equiv\mu-V(\bfr_i)$ with $\mu$ the global chemical potential and $V(\bfr_i)= V_{\rm tr}\times(x_i^2 + y_i^2)$ being the isotropic trapping potential at lattice position, $\bfr_i=(x_i,y_i)$. Within a 5\% uncertainty in the total particle number $N$, direct comparisons of the density images show excellent and quantitative agreements (see Fig. \ref{CCexp}(a) and (b)) both in the weak and strong lattice strengths. A small deviation in the trap center of Fig. \ref{CCexp}(b) is due to some experimental equilibration issues [\oncite{Chin_transport}]. Hence, our results convincingly indicate that the bosonic atoms loaded into an optical lattice is an excellent quantum simulator. 

\begin{figure}
\includegraphics[width=8.5cm]{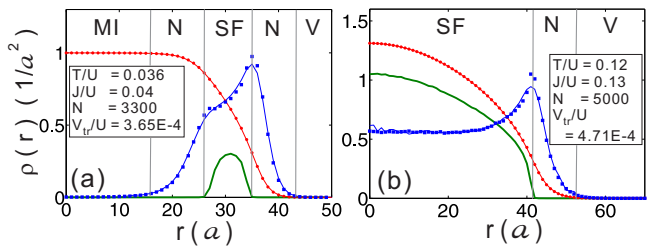}
\caption{(Color online)(a) and (b) are typical density profiles (red lines) and local compressibility (blue lines, arbitrary unit) in a trap system, obtained from the QMC calculation. Red dots and blue squares are results obtained by LDA, almost identical to the QMC results. The thick green lines shows the local superfluid density within the LDA.  Here $N(V)$ stands for normal phase(vacuum).
}
\label{PDLDA}
\end{figure}

In Fig. \ref{CCexp}(c), we show the finite temperature phase diagram in a uniform 2D system. The boundaries between SF and normal phases (red lines) are determined by vanishing SF density[\oncite{WORMWINDING}]. Strictly speaking, MI phase doesnot exist at $T>0$, but we can still definite its regime (blue lines) empirically by a small compressibility, $\kappa\equiv\partial n/\partial \mu<0.02/U$. As a result, one can clearly see how both the SF and MI regimes shrink as $T$ increases, leading to the normal phase in the intermediate regime. Such finite temperature phase diagram of a uniform system can help in the determination of the phase boundaries in a trapped system within the LDA [\oncite{Zhou_Qi_3D}], as shown in Figs. \ref{PDLDA}(a) and (b). The density profile ($\rho(\bfr)$) as well as local compressibility ($\kappa(\bfr)\equiv\partial \rho(\bfr)/\partial\mu$) are obtained within the QMC simulation, but both of them can be well-reproduced within the LDA, except for the narrow regime near the phase boundary [\oncite{Lode_Qi}]. Such rounding of $\kappa(\bfr)$ originate from the finite size effect and the finite temperature Berezinskii-Kosterlitz-Thouless (BKT) transition. From our direct comparison, however, the phase boundary determined by the largest curvature points of $\kappa(\bfr)$ is still very close to the position obtained within the LDA (within one or two lattice sites), and also agrees with the onset of the local SF density. Therefore, practically speaking, one can still apply LDA to estimate the phase boundary even in the 2D trapped system at finite temperature. In the following, we will use another scheme to investigate the phase boundary as well as the quantum criticality near the SF-MI quantum phase transition from the density profile.

It is well-known that near the second order phase transition, the correlation length diverges and universal scaling behaviors are expected in many physical quantities. In the context of ultracold atoms, the universal scaling behaviors of a classical superfluid phase transition have been observed in a 3D [\oncite{scaling_Esslinger}] and 2D [\oncite{scaling_Cheng}] weakly interacting Bose gas when the temperature is near the transition temperature. However, the scaling behavior near the quantum (zero temperature) phase transition point is still unexplored in current cold atom experiments. To provide useful indications for future experiments, we shall examine the finite-temperature calculation in quantitative detail, as proposed in Ref. [\oncite{Jason,Mueller}]. In the low-energy limit,
it is well known that the SF-MI transition can be described by one of the following two effective models [\oncite{Sachdev,Mueller}]: the dilute Bose gas model (DBG) for the phase transition in the edges of the MI lobe (for example, along the $C_1-C_2$ line of Fig. \ref{CCexp}(c)), and the $O(2)$ rotor model (or $XY$-model) near the tip of MI lobe, where the particle-hole symmetry holds. 

Near the transition point, the free energy becomes singular in an infinite system, and its singular part, $f_s$, should be a just function of $\mu$ and $T$ for a given $J/U$. As a result, a universal scaling theory gives $f_s(\mu,T) \sim T^{\frac{d}{z}+1} F((\mu-\mu_c)/T^{1/z\nu})$ [\oncite{Jason,Fisher}], where $d$ is dimensionality, $z$ is the dynamical exponent, $\nu$ is the correlation length exponent, $\mu_c$ is the chemical potential at zero temperature phase transition point, and $F$ is the universal scaling function. As a result, we can obtain the singular part of the particle density, $n_s=n-n_r=\partial f_s/\partial\mu$, and compressibility, $\kappa_s=\kappa-\kappa_r=\partial n_s/\partial\mu$, as following (here $n_r(\kappa_r)$ are the regular part of density (compressibility)):
\be
n_s(\mu, T) &\sim & T^{\frac{d}{z}+1-\frac{1}{z\nu}} G\left(\frac{\mu-\mu_c}{T^{1/z\nu}}\right),
\label{n_s}
\ee
\be
\kappa_s(\mu, T) &\sim & T^{\frac{d}{z}+1-\frac{2}{z\nu}} H\left(\frac{\mu-\mu_c}{T^{1/z\nu}}\right).
\label{k_s}
\ee
where $G(x)\equiv dF(x)/dx$ and $H(x)=dG(x)/dx$ are also universal functions. Combining them we could eliminate $\mu-\mu_c$ and obtain 
\be
\kappa_s(n_s,T)&=&T^{\frac{d}{z}+1-\frac{2}{z\nu}}H\left(G^{-1}\left(n_s T^{\frac{1}{z\nu}-1-\frac{d}{z}}\right)\right),
\label{ks_ns}
\ee
where both $n_s$ and $\kappa_s$ can be measured easily [\oncite{Chin}].

\begin{figure}
\includegraphics[width=8.5cm]{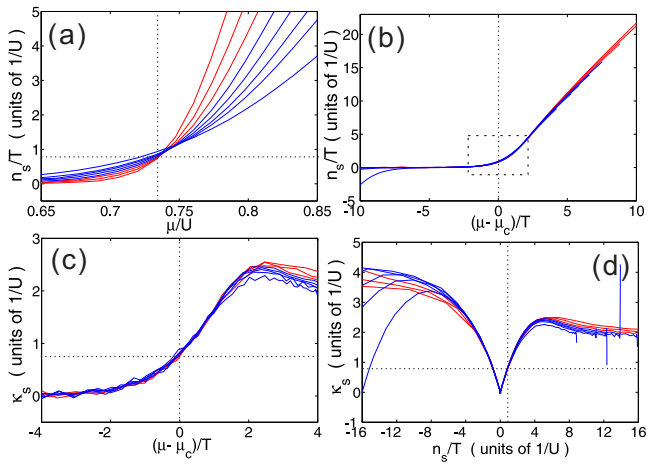}
\caption{(Color online) 
(a) $n_s/T$ as a function of $\mu(r)$ in a trapped system for various temperature. (b)-(d) are respectively the universal scaling function $G$, $H$, and $H(G^{-1})$ as defines in Eqs. (\ref{n_s})-(\ref{ks_ns}). Red(blue) lines are taken from the low(high) temperature data, $0.02<T/U<0.03$($0.03<T/U<0.0625$). The vertical and horizontal dotted lines indicate the critical values ($C_1$ point of Fig. \ref{CCexp}(c)). The rectangular window shows the parameter regime of (a). The sharp noise in (d) results from the flat density at the cloud center. Here we use $J/U=0.03$, $V_{\rm tr}/U=1.75\times 10^{-4}$, $N\sim 2\times 10^4$, and exponents of the 2D DBG model here.}
\label{scaling}
\end{figure}

In the following analysis, we will first concentrate on the generic case, i.e the class of DBG model, and then show results of the $O(2)$ rotor model later. It is well-known that $(z,\nu)=(2,1/2)$ and $(z,\nu)=(1,1)$ respectively for them when fixing $J/U$ [\oncite{Mueller,Fisher}]. According to Ref. [\oncite{Jason}], the cross point of $n_s /T$ at different temperature can give the value of $\mu_c$ from Eq. (\ref{n_s}), as clearly observed in our simulation data shown in Fig. \ref{scaling}(a). Since we consider the regime with $n_r=1$ and $n_s>0$, it gives us the transition point $C_1$ in Fig. \ref{CCexp}(c). Similarly we can also obtain the position of $C_2$ (not shown), so that their difference gives the zero temperature gap for a given $J/U$. We also have performed the finite size scaling in a trapped system (proposed in Ref. [\oncite{Jason,trap_scaling}]) to identify the precise value of $\mu_c$. The obtained result is identical ($<0.1\%$) to the result of a uniform system after finite size scaling [\oncite{scaling_note}], and also very close ($<1\%$) to the one we obtain above in a trap system (Fig. \ref{scaling}(a)). Our results varify the proposal to map out the zero temperature phase diagram in a realistic trapped system.

After determination of $\mu_c$, in Fig. \ref{scaling}(b) we show the rescaled singular density ($n_s/T$) as a function of rescaled chemical potential, $(\mu-\mu_c)/T$, in the DBG model, and find a very nice convergence for the universal scaling function $G$ (see Eq. (\ref{n_s})) in a wider temperature regime. This is a clear evidence of the scaling theory near the critical point of SF-MI transition. Similarly, we also show the universal functions for $H$ and $H(G^{-1})$ in Figs. \ref{scaling}(c) and (d) respectively. The later results are also shown for both the particle side and hole side of the phase transition points (i.e. the $C_1$ and $C_2$ points in Fig. \ref{CCexp}{c)).

In our calculation here, we find that the universal scaling function is very sensitive to the value of exponents, i.e. no intersection point for $\mu_c$ and no convergence of the scaling function can be obtained if using exponents of other universal classes. This can be also observed when we calculate the universal scaling function near the Mott tip, as shown in Fig \ref{ZEROT}(a) and (b). For these data at $J/U=0.054$ (i.e. $C_3-C_4$ line in Fig. \ref{CCexp}(c)), exponents obtained from the $O(2)$ rotor model works much better than the DBG model, consistent with the general understanding that the low energy effective model at the Mott tip is governed by the former. The crossover from the DBG model to the $O(2)$ rotor model near the Mott tip is of great interest, but is beyond the scope of this paper.
 
\begin{figure}
\includegraphics[width=8.5cm]{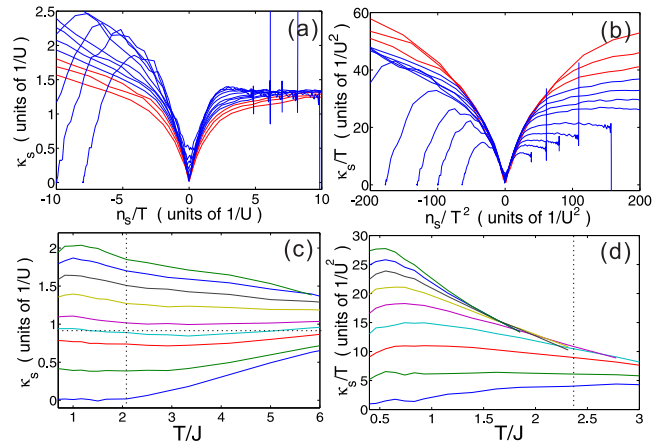}
\caption{(Color online) (a) and (b) are scaling function $H(G^{-1})$ near the tip of MI lobe ($J/U=0.054$), where the scaling exponents are from the DBG model and the $O(2)$ model respectively. Red(blue) lines are for the lower(higher) temperature as indicated in Fig. \ref{scaling}. (c) shows how $\kappa_s$ changes as a function of temperature for several typical values of $n_s/T$ at $J/U=0.03$ (same parameters as Fig. \ref{scaling} in the DBG model). The vertical dashed line indicates the maximum temperature ($T/U=0.0625$) used in Fig. \ref{scaling}, and the horizontal dotted line indicates $\kappa_c$ at the critical point. (d) Similar to (c) but for $J/U=0.054$ in the $O(2)$ rotor model.}
\label{ZEROT}
\end{figure}
From the experimental point of view, however, it is somehow more important to know the characteristic temperature, $T^\ast$, below which the universal scaling behavior is observable. The effective model itself cannot provide this information, and therefore our {\it ab-initio} QMC data becomes very crucial. In Fig. \ref{ZEROT}(c), we show how the universal scaling of the DBG model is affected by temperature at $J/U=0.03$. We can see that $\kappa_s$ keeps almost a constant in the low temperature limit up to some higher temperature, $T^\ast$, above which their universal value starts to deviate due to higher order effects. For compressibility near $\kappa_c$ (i.e. $\mu\sim\mu_c$), such $T^\ast$ becomes much larger. If we set the minimum range for observing the scaling law in real space (within LDA) is about 10 lattice sites, the corresponding $T^\ast\sim 6J$ is about the order of the single particle band width. Although such low temperature measurement is challenging, but should be still experimentally accessible in the near future. When $J/U$ is close to the MI tip, due to the crossover from the DBG model to the $O(2)$ model, we find that $T^\ast$ is reduced and the range to observe the universal scaling behavior becomes smaller (see Fig. \ref{ZEROT}(d)). Therefore, at least within the QMC simulation, our results confirm the possibility of observe the universal scaling theory near the critical point in the present experimental parameter regime. 

\begin{figure}
\includegraphics[width=8.5cm]{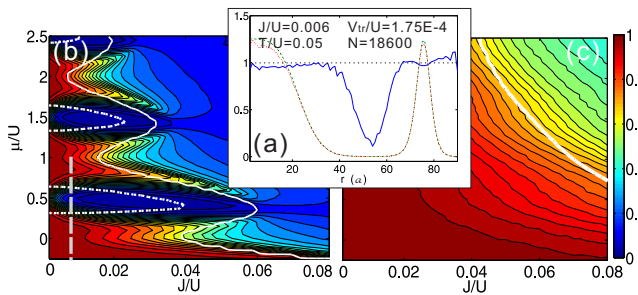}
\caption{(Color online) (a) shows the calculated values of $k_BT\kappa(\bfr_i)$ (red dashed line) and $\langle\delta\hat{n}_i^2\rangle$ (green dotted line) and their ratio ($\chi(\bfr_i)$, blue solid line) for a typical trapped system. (b) and (c) show the contour plot of $\chi$ as a function of $J/U$ and $\mu$ in a uniform 2D system at $T/U=0.05$ and 0.25 respectively. The solid and dashed lines in (a) indicate the SF and MI boundaries. The vertical line in (b) is the range of chemical potential in (a).}
\label{FDT}
\end{figure}

Accurate thermometry is the essence in the observation of universal scaling behavior. Earlier scheme to fit the outskirt density profile with a noninteracting model hardly works in the strongly interacting regime we consider here. 
It has been proposed that the fluctuation-dissipation theorem (FDT) can be a feasible thermometry scheme (independent of any theoretical model) [\oncite{JasonTM,HIGHTexpansionTAMA}], but in realistic experiments, it is more practical to consider the local version of the FDT. We start from the identity, 
$\frac{\partial N}{\partial \mu} = \langle \delta \hat{N}^2\rangle/k_BT$, where $\hat{N}\equiv\sum_i\hat{n}_i$ is the total number operator. Therefore, the local compressibility can be related to the local density fluctuation by:
$\kappa(\bfr_i)\equiv\frac{\partial n_i}{\partial \mu}=\langle \delta \hat{n}_i^2\rangle/k_B T + \sum_{j\neq i} \langle \delta \hat{n}_i \delta \hat{n}_j\rangle/k_BT$,
where the last term is the non-local density correlations. In the normal phase regime away from the critical point, this term should be negligible due to the short correlation length, i.e. $k_BT \kappa(\bfr_i)=\langle \delta \hat{n}_i^2\rangle$. In order to investigate the regime when the local FDT is valid, we define $\chi(\bfr_i)\equiv k_BT\kappa(\bfr_i)/\langle\delta \hat{n}_i^2\rangle$, and show the result of a typical parameter in Fig. \ref{FDT}(a), where there is a wide regime (not only in the outskirt regime) to find a constant plateau, which can estimate the system temperature with a reasonably high precision. More generally, in Fig. \ref{FDT}(b) and (c), we find the regime for $\chi\approx 1$ is alway in the normal phase regime of small $J/U$ (and away from the Mott tip), and then occupies the entire lower-left corner when the temperature is higher. Our results show that even is a pretty low temperature (but still $T>J$ as in the present experimental limit), the wide range of normal phase helps the justification of the local FDT as a thermometry. This method takes the advantage that only local data measurement is needed, and can be also used to study the non-local correlation effects in the current experiments. When deep inside the SF regime (more particle number per site) or when $T<J$, non-local correlations become significant and the local FDT should be corrected with higher order terms as suggested in Ref. [\oncite{HIGHTexpansionTAMA}].

In conclusion, we show the direct comparison between the {\it in-situ} experimental data and the {\it ab-initio} numerical calculation for bosonic atoms inside a two-dimensional optical lattice with a trapping potential. We map out the quantum phase diagrams in a uniform system by using the knowledge of the density profile only. Besides, we note that the universal scaling function as well as other quantum critical properties can be observed if the system temperature is lower than the single particle band width. Finally we investigate the regime when the LDA and the local FDT works, where one can use them to extract phase diagram and system temperature. Our results can be directly applied to the current experimental and for future investigation of the quantum simulator.

We appreciate the discussion with C. Chin, N. Gemelke, T.-L. Ho, C.-L. Hung, E. Mueller, L. Pollet, and Q. Zhou, especially Chicago's group for experimental data and L. Pollet's QMC code for benchmark. We also thank R.-K. Lee for the computation resources. This work is supported by NSC(Taiwan).



\begin{thebibliography}{99}
\bibitem{scaling_note}
Considering the finite size ($L$) and finite temperature ($T$) effects, the superfluid density ($\rho_s$) in a uniform system should have the following scaling form: $\rho_s(\delta,T,L)\sim\delta^{\nu(d+z-2)}R(L\delta^\nu,T/\delta^{z\nu})$, where $\delta=\mu-\mu_c$. We then determine the phase boundary for various sizes at different temperature. The obtained $\mu_c$ is almost the same ($<0.1\%$) as the result calculated from the trap scaling method [\oncite{trap_scaling}].
\end{thebibliography}
\end{document}